\documentclass[fontsize=10pt]{llncs}
\usepackage{latexsym, amssymb, amsmath, amsfonts}

\usepackage{amsthm}
\usepackage{stmaryrd}
\usepackage{semantic}
\usepackage[all]{xy}
\usepackage{dsfont}
\usepackage{enumerate}
\usepackage{stackengine}
\usepackage{mathtools}
\usepackage{pgfplots}
\usepackage{listings}
\usepackage{makecell}  
\usepackage{paralist}
\usepackage{pdfpages}
\usepackage{mathrsfs} 
\usepackage{scalerel,accsupp} 
\usepackage{xstring}
\usepackage{verbatim}

\newcommand{\draftnote}[2][{$\mbox{}^\star$}]{#1 \marginpar
{\small
\raggedright
\textsf{\hspace*{-1.6ex}#2} 
}}

\newcommand{\signedcomment}[3]
  {\draftnote[$\mbox{}^{\color{#2} \star}$]{{\color{#2} \small [#1] #3}} }

\newcommand{\protectedTable}{\setName{ProtectedTable}}

\newcommand{\mySim}[3] { \relation{#1} \sim \relation{#2} }

\newcommand{\eqdef}
 {\mathrel{\stackrel{{\scriptscriptstyle\mathrm{def}}}{=}}}

\DeclareMathAlphabet{\mathsc}{OT1}{cmr}{m}{sc}

\newcommand{\proofcase}[2][Case]{\noindent  
 \raisebox{2ex}{\mbox{}} 
 \textbf{#1: #2}.~~}

\newcommand{\ForAll}[2]
           {\forall #1 .\,#2}

\newcommand{\Nat}{\mathbb{N}}
\newcommand{\Real}{\mathbb{R}}

\newcommand{\RangeS}{S}

\newcommand{\Set}[1]{ \{ #1 \} }
\newcommand{\SetDef}[2]{ \{ #1 \; | \; #2 \} }
\newcommand{\Union}{\cup}

\newcommand{\relation}[1]{#1}

\newcommand{\opSymbol}[1]{{\mathsf{#1}}} 
\newcommand{\opName}[1]{\opSymbol{#1}}
\newcommand{\setName}[1]{\mathrm{\mathbf{#1}}}
\newcommand{\varName}[1]{\mathit{#1}}

\newcommand{\codomain}{\opName{codomain}} 
\newcommand{\Probability}[1] {\Pr[#1]}

\newcommand{\upto}{,\!\makebox[1em][c]{.\hfil.\hfil.},} 

\usepackage{esvect} 
\newcommand{\vect}[1] {\vv{#1}}

\newcommand*{\llbrace}{%
  \BeginAccSupp{method=hex,unicode,ActualText=2983}%
    \textnormal{\usefont{OMS}{lmr}{m}{n}\char102}%
    \mathchoice{\mkern-4.05mu}{\mkern-4.05mu}{\mkern-4.3mu}{\mkern-4.8mu}%
    \textnormal{\usefont{OMS}{lmr}{m}{n}\char106}%
  \EndAccSupp{}%
}
\newcommand*{\rrbrace}{%
  \BeginAccSupp{method=hex,unicode,ActualText=2984}%
    \textnormal{\usefont{OMS}{lmr}{m}{n}\char106}%
    \mathchoice{\mkern-4.05mu}{\mkern-4.05mu}{\mkern-4.3mu}{\mkern-4.8mu}%
    \textnormal{\usefont{OMS}{lmr}{m}{n}\char103}%
  \EndAccSupp{}%
}

\newsavebox{\lXbrace}
\savebox{\lXbrace}{$\llbrace$}
\newsavebox{\rXbrace}
\savebox{\rXbrace}{$\rrbrace$}

\def\lxbrace{%
   \hstretch{0.6}{\scalerel*{\usebox{\lXbrace}}{\llbrace}}}
\def\rxbrace{%
   \hstretch{0.6}{\scalerel*{\usebox{\rXbrace}}{\rrbrace}}}

\makeatletter
\newsavebox{\mystrut}
\newcommand{\mSetDefBig}[3][1.5ex]
{
\dimen@ #1
\savebox{\mystrut}{\raisebox{-1\dimen@}{\rule{0pt}{3\dimen@}}}
\scalerel*{\lxbrace}{\usebox{\mystrut}}
  #2 \; \scalerel*{\vert}{\usebox{\mystrut}} \; #3\,
   \scalerel*{\rxbrace}{\usebox{\mystrut}}
}
\makeatother

\newcommand{\apply}[2] {#1(#2)} 

\newcommand{\Configuration}[3] {\langle #1, #2, #3 \rangle}
\newcommand{\Conf} {\mathds{C}}
\newcommand{\InitConf} {\setName{Init}}
\newcommand{\Configs} {\setName{Config}} 

\newcommand{\Silent} {\tau}
\newcommand{\Budget} {B}
\newcommand{\Env}[1] {#1}

\newcommand{\Program}[1] {#1}
\newcommand{\AllPrograms} {\mathds{P}}

\newcommand{\Query}[1] {{Q_{#1}} }
\newcommand{\QueryVec}[1]{\vect{Q}, #1}  

\newcommand{\QueryAct}{\opName{query}} 
\newcommand{\QuerySet}{\setName{Query}} 

\newcommand{\RecordSet}{\setName{Record}}

\newcommand{\Length}[1] {|#1|}
\newcommand{\Size}[1] {\opName{size}(#1)}

\newcommand{\rSim}[1]{\mathrel{\overset{#1}{\sim}}}

\newcommand{\Mstep}[2] { \xRightarrow{#1}_{#2} } 
\newcommand{\Sstep}[2] { \xrightarrow{#1}_{#2} } 
\newcommand{\Pstep}[1] { \xrightarrow{#1} } 

\newcommand{\EmptyTrace} {[]}

\renewcommand{\epsilon}{\varepsilon} 

\newcommand{\trace} {t}

\newcommand{\TVar}{\setName{TVar}}
\newcommand{\tvar} {\varName{tv}}

\newcommand{\Val}{\setName{Val}}

\newcommand{\AllTables} {\setName{Table}}
\newcommand{\SensitivityFunc} {\opName{stability}}

\newcommand{\ProgAct} {\setName{ProgAct}} 
\newcommand{\Act} {\setName{Act}}         

\definecolor{dkgreen}{rgb}{0,0.6,0}
\definecolor{gray}{rgb}{0.5,0.5,0.5}
\definecolor{mauve}{rgb}{0.58,0,0.82}
\definecolor{light-gray}{gray}{0.25} 

\lstdefinestyle{csharp} {
aboveskip=3mm,
belowskip=3mm,
showstringspaces=false,
columns=flexible,
basicstyle={\footnotesize\ttfamily},
numberstyle={\tiny},
numbers=left,
keywordstyle=\color{blue},
commentstyle=\color{dkgreen},
stringstyle=\color{mauve},
breaklines=true,
breakatwhitespace=true,
tabsize=3,
morecomment = [l]{//},
morecomment = [l]{///},
morecomment = [s]{/*}{*/},
morestring=[b]",
sensitive = true,
morekeywords = {async, await, abstract,
event, new, struct,
as, explicit, null, switch,
base, extern, object, this,
bool, false, operator, throw,
break, finally, out, true,
byte, fixed, override, try,
case, float, params, typeof,
catch, for, private, uint,
char, foreach, protected, ulong,
checked, goto, public, unchecked,
class, if, readonly, unsafe,
const, implicit, ref, ushort,
continue, in, return, using,
decimal, int, sbyte, virtual,
default, interface, sealed, volatile,
delegate, internal, short, void,
do, is, sizeof, while,
double, lock, stackalloc,
else, long, static,
enum, namespace, string }
}

\usepackage[
            compact
           ]{titlesecMod} 
\titlespacing{\section}{0pt}{*2}{*1} 
   
\titleformat{\paragraph}[runin]
   {\normalfont\normalsize\bfseries}{\theparagraph}{1em}{}

\AtBeginDocument{%
\addtolength\abovedisplayskip{-0.4\baselineskip}%
\addtolength\belowdisplayskip{-0.4\baselineskip}%
\addtolength{\textfloatsep}{-4ex}
\addtolength{\intextsep}{-4ex}
\addtolength{\abovecaptionskip}{-2ex}
\addtolength{\belowcaptionskip}{2ex}
}

\renewcommand\floatpagefraction{.9}
\renewcommand\topfraction{.9}

\usepackage{url}
\urldef{\mailsa}\path|{hamide,dave}@chalmers.se|   
\newcommand{\keywords}[1]{\par\addvspace\baselineskip
\noindent\keywordname\enspace\ignorespaces#1}

\begin{document}

\lstdefinestyle{ProPercode}{
    basicstyle=\scriptsize\ttfamily,
    breaklines=true,
    breakatwhitespace=true,
    tabsize=1,
    resetmargins=true,
    xleftmargin=0pt,
    frame=none
} 


\mainmatter  

\title{Featherweight PINQ}

\titlerunning{Featherweight PINQ}

%
%
\author{Hamid Ebadi%
\and David Sands}
\authorrunning{Featherweight PINQ}

\institute{Department of Computer Science and Engineering, \\
          Chalmers University of Technology \\
          Gothenburg, Sweden \\
\mailsa}

%
%

\maketitle

\begin{abstract}

Differentially private mechanisms enjoy a variety of composition properties. Leveraging these, McSherry introduced PINQ (SIGMOD 2009), a system empowering non-experts to 
construct new differentially private analyses. PINQ is an LINQ-like API which provides automatic privacy guarantees for all programs which use it to mediate sensitive data manipulation.   In this work we introduce \emph{featherweight PINQ}, a formal model capturing the essence of PINQ.  We prove that any program interacting with featherweight PINQ's API is differentially private.

%
%
%


\end{abstract}

\section{Introduction}

Differential privacy 
\cite{Dwork06dp,dwork2008survey,Dwork:Firm} shows that by 
adding the right amount of noise to statistical queries, one can get useful results, and at the same time provide a quantifiable notion of privacy. The definition of differential privacy for a query mechanism (a randomized algorithm) is made by comparing the results of a query on any database with or without any one individual: a query $Q$ is $\epsilon$-differentially private if the difference in probability of any query outcome on a data-set only changes by a factor of $e^\epsilon$ (approximately $1+\epsilon$ for small $\epsilon$) whenever an individual is added or removed. 

Of the many of papers on differential privacy, a mere handful (at the time of writing) describe implemented systems which provide more than just a static collection of differentially private operations. 
The first such system is the PINQ system of McSherry \cite{mcsherry2009privacy}. 
PINQ is  
designed to allow non-experts in differential-privacy to build privacy-preserving data analyses.  The system works by leveraging a fixed collection of differentially private data aggregation functions (counts, averages, etc.), and a collection of data manipulation operations, all embedded with a LINQ-like interface from otherwise arbitrary C\# code.  PINQ mediates all accesses to sensitive data in order to keep track of the sensitivity 
 of various computed objects, and to ensure that the intended privacy budget $\epsilon$ is not exceeded; a budget could be exceeded by answering  too many queries with too high accuracy. 
In this way PINQ is intended to make sure that the analyst (programmer) does not inadvertently break differential privacy. 

\paragraph{Foundations of PINQ}
McSherry argues the correctness of PINQ by pointing out the foundations upon which PINQ rests. In essence these are: 
\begin{enumerate}
\item A predefined collection of aggregation operations (queries) on tables, each with a parameter specifying the required degree of differential privacy. Standard aggregation operations such as (noisy) count and average are implemented. The core assumption is that each aggregation operation $Q$ with noise parameter $\epsilon$, written here as $Q_\epsilon$, is an $\epsilon$-differentially private randomised function. 

\item Sequential composition principle: if two queries performed in sequence (e.g. with differential privacy $\epsilon_1$ and $\epsilon_2$ respectively) then the overall level of differential privacy is safely estimated by summing the privacy costs of the individual queries  ($\epsilon_1 + \epsilon_2$). 

\item Parallel composition principle: if the data is partitioned into disjoint parts, and a different query 
is applied to each partition, then the overall level of differential privacy is safely estimated by taking the maximum of the costs of the individual queries. 

\item Stability composition: the stability of a database transformation $T$ is defined to be $c$ if when ever you add $n$ extra elements to the argument of $T$, the result of $T$ changes by no more than $n \times c$ elements. If you first transform a database by $T$, then query the result with an $\epsilon$-private query, the privacy afforded by the composition of the two operations is safely approximated by $c \times \epsilon$.
\end{enumerate}


These "foundations" of PINQ provide an intuition about how and why PINQ works, but although a novel aim of PINQ was ``providing formal end-to-end differential privacy guarantees
under arbitrary use'', the foundations 
are inadequate to build an end-to-end correctness argument 
since they fall short of describing number of PINQ features of potential relevance to the question of its differential privacy:
\begin{itemize}
\item Sequential composition is described in an oversimplified way, assuming that the queries 
are chosen independently from each other. In practice the second query of a
  sequence is issued by client code which has received the result of
  the first query. Thus the second query may depend 
on the outcome of
  the first. To argue correctness this adaptiveness should be
  modelled explicitly. 

\item Parallel queries partition data, but the data which is
  partitioned might not be the original input, but some intermediate
  table. The informal argument for taking the
  maximum of the privacy costs of the query on each partition relies
  on the respective queries applying to disjoint data points. 
But the data might not be disjoint when seen from the perspective of the original data set of individuals. Data derived from a participant might end up in more than one partition, so a correctness argument
  must model this possibility to show that it is safe.

\item As for sequential composition, parallel queries are not
  parallel at all, but can be adaptive - the result of a query on one
  partition might depend on the result of a query on another.  This
  means that the implementation greatly complicates the bookkeeping
  necessary to track the "maximum" cost of the queries.

\item The foundations suggest how to compute the privacy cost of
  composed operations from their privacy and stability
  properties. But in practice PINQ does not \emph{measure} the amount of privacy lost by a PINQ program, it \emph{enforces} a stated bound. 
  Because of this, there are two kinds of results from a query: the normal noisy answer, or an exception. An exception is thrown if answering the query normally would break the global privacy budget.  To prove differential privacy it is not enough that the query is differentially private in the normal case -- it must also be shown to be private in the case when an exception is thrown, since this information is communicated to the program.
\end{itemize}
In this paper we provide a foundation for PINQ by defining a minimalistic semantics, 
\emph{Featherweight PINQ}, intended to model it's essence, while at the same time abstracting away from less relevant implementation details. By idealising the interface we make clear the intended implementation, but not the details of its realisation in any particular language. 
Thus we model the client program completely abstractly as a deterministic labelled transition system which interacts with tables via the PINQ-like API but which is otherwise unconstrained.
For this model we instantiate  the definition of differential privacy, taking into account the interactive nature of the system, and prove that  Featherweight PINQ provides differential privacy for any program. 


 




\section{PINQ} 	

In this section we provide a brief description of the PINQ system from
the user perpective. PINQ is a .NET API which provides an interface similar to the Language
Integrated Queries (LINQ) that is a language extension to .NET. 
Analyses that use PINQ are typically written in C\#.

Listing~\ref{pinqcode} shows a code fragment for a sample analysis producing the average ages of adult males and adult females, respectively and then separately compute the average of age for all individuals. 

\lstset{escapechar=@,style=csharp}
\begin{lstlisting}[caption=PINQ sample code, label=pinqcode]
var agent = new PINQAgentBudget(budget);
var data = new PINQueryable<recordstype>(rawdata, agent);
var adults = data.Where(x => x.age > 17);
var genders = new [] {0,1};
var parts = adults.Partition(genders, x=>x.gender);
foreach (var a in genders) {
   Console.WriteLine("Average age of {0} is {1}",
                       a==0? "Males " : "Females ",
                       parts[a].NoisyAverage(budget/2, x=>x.age)
}
Console.WriteLine("Average age (all):" + data.NoisyAverage(c))
\end{lstlisting}
The first two lines of the program initialises a PINQueryable object with sample sensitive data (\texttt{rawdata}) structures and set the privacy limit (\texttt{budget}). A PINQueryable object is a wrapper to the database which enables PINQ to track the properties that are relevant for differential privacy. The supplied ``agent'' parameter expresses the 
amount of differential privacy that the system will enforce on this database. 

The analysis starts by selecting (line 3) a subset of records of interest (those who are adults). Behind the scenes PINQ records the fact that the \emph{stability} of \verb!data! is unchanged: adding a single record to the \verb!rawdata! does not change the size of the result of this transformation by more than a single record. 

In line 5 a partitioning operation splits the data into two groups based on the gender field (0 for Male, 1 for Female). Partition is not a standard LINQ/SQL style operation, but is specific to PINQ. 
For each partition (i.e. for each gender), 
the code outputs a noisy average of the \verb!age!. \verb!NoisyAverage! is one of a collection of built-in differentially private primitive aggregation operations provided by PINQ. The amount of differential privacy for each query in the loop is \verb!budget/2!.  After executing the \verb!foreach! loop there will be \verb!budget/2! of the original budget remaining. The outcome of the last line depends on the accuracy/privacy parameter \verb!c!. If $c$ is larger than \verb!budget/2! the program will throw an exception (because answering the query with that degree of precision would break the budget).



\section{Idealised Program}

In this section we describe the abstract model of the program and API to the PINQ operations. In the section thereafter we go on to model the PINQ internals, what we call the \emph{protected system},  before combining these components into a the overall model of Featherweight PINQ.



The first thing that we will abstract away from is the host programming language. Here one could chose to model a simple programming language, but it is not necessary to be that concrete.  Instead we model a program as an arbitrary deterministic system that maintains its own internal state, and issues commands to the PINQ internals.  In this sense we idealise PINQ by assuming that the API cannot be bypassed.  In fact the PINQ system does not successfully encapsulate the all the protected parts of the system, and so some programs can violate differential privacy by bypassing the encapsulation \cite{haeberlen2011differential}, or by using side effects in places where side-effects are not intended. By idealising the interface we make clear the intended implementation, but not the details of its realisation in any particular language. 
By treating programs abstractly we also simplify other features of PINQ including aspects of its architecture which promote certain forms of extensibility.






Before describing the program model it is appropriate to say a few words about the \emph{protected system} (described formally in the next section). The protected system contains all the datasets (tables) manipulated by the program.  Since these are the privacy sensitive data, we only permit the program to access them via the API.  The protected system tracks the stability of all the tables which it maintains, together with a global budget. Our program interacts with the protected system by the following operations:


\paragraph{Assignment}
Tables in the protected system are referred to via \emph{table variables}. A program can issue an assignment command.
The model allows the program to manipulate a table using transformation that assign a new value to table variables.

The general form of assignment is of the form  $ \tvar := F(\tvar_1,\ldots,\tvar_n)$, where $F$ is taken from a set of \emph{function identifiers} representing a family of transformations with bounded stability (i.e. for each argument position $i$ there is a natural number $c_i$ such that if the size of the $i$th argument changes by $n$ elements, then the result will change by at most $c_i \cdot n$  elements). This stability requirement comes from PINQ and is discussed in more detail in the next section. Transformations include standard operations such as the \verb!.Where(x => x.age > 17)! from the example in listing~\ref{pinqcode}, and  simple assignments $t_1 := t_2$ (taking $F$ to be the identity function), as well as assignments of literal tables (the case when $F$ has arity $0$).




\paragraph{Query}
The only other operation of the PINQ API is the application of a primitive differentially private query.
In the example above we saw a compound transformation and query operation
\verb!parts[a].NoisyAverage(budget/2, x=>x.age)!. It is sufficient to model just the query, since the transformation (\verb!x=>x.age!) can be implemented via an intermediate assignment. Thus we assume a set of primitive queries $\QuerySet$, ranged over by $Q$, which take as argument a positive real (the $\epsilon$ parameter) and a table, and produce a discrete probability distribution over a domain of result values $\Val$.  


We generalise the single query operation to a \emph{parallel query}, with syntax 
$\QueryAct(\tvar,f,\vect{Q},\epsilon)$, where 
%
%
%
%
%
\begin{enumerate}
    \item $\tvar$ is the table variable referring to the table that will be used for the analysis,
    \item $f$ is the partitioning function that maps each record to an index in $\codomain(f)= \Set{1,\ldots, k}$ for some $k \in \Nat$,   
    \item $\vect{Q}$ is a vector of $k$ 
queries from $\QuerySet$.
\end{enumerate}
The execution of this operation (as described in the next section) involves computing the sequence of randomised values 
\[ Q_i(\epsilon,\SetDef{r \in T}{f(r) = i}), i \in \codomain(f) \]
where $T$ is the table bound to $\tvar$.
This is the ``parallel query'' operation described informally in the description of PINQ \cite{mcsherry2009privacy}. We use a single $\epsilon$ for all queries because if we chose an $\epsilon_i$ for each query the privacy cost will be maximum of all the epsilons in any case, so we may as well enjoy the accuracy of the largest epsilon.  However, we note that the implementation of PINQ is more general than this, since the queries on each partition may be performed in an adaptive way. Here we are making a trade-off in keeping our model simple at the expense of not proving differential privacy for quite as general a system.

\paragraph{Client Program Model}
The above abstraction of the PINQ API allows us to abstract away from all internal details of the programming language using the API. Following  \cite{PersonalisedDifferentialPrivacy} we model a program as an arbitrary labelled transition system with labels representing the API calls:
\begin{definition}[$\ProgAct$ Labels]\label{def:progact}
The set of program action labels $\ProgAct$, ranged over by $a$ and $b$, are defined as the union of three syntactic forms:
\begin{enumerate}
\item the distinguished action $\tau$, representing computational progress without interaction with the protected system, 
\item $tvar := F(\tvar_1 , \ldots , \tvar_n)$ where $F$ is a function identifier, i.e. the formal name of a transformation operation of arity $n$, 
\item $\QueryAct(\tvar,f,\vect{Q},\epsilon)?\vect{v}$, where $f$ is a function from records to $\Set{1,\ldots, k}$ for some $k > 0$, where $\vect{v}$ is a vector of values in $\Val^k$, and $\vect{Q}$ is a vector of $k$ queries.
\end{enumerate}
\end{definition}



Every label represents an interaction between a client program and the protected system. The labels represent the observable output of a system which are a sequence of those actions: internal (silent) steps ($\Silent$) modelling no interaction, and vectors of values $\vect{v}$ which are the results of some query being answered and returned to the program. 

To define these transitions, we assume a client program modelled by a labelled transition system modelling the API to the protected system.
For client programs, the label corresponding to a query call is of the form $\QueryAct(\tvar, f, \QueryVec{\epsilon})\ ?\ \vect{v}$, 
and models the pair of query and the returned result (as described before) as a single event. This allows us to model value passing with no need to introduce any specific syntax for programs. Note that the value returned by the query is known to the program, and the program can act on it accordingly. From the perspective of the program and the protected system together, this value will be considered an observable output of the whole system.

\begin{definition}[Client Program]\label{def:program}
A \emph{client program} is a labelled transition system $\langle \AllPrograms, {\rightarrow}, P_0 \rangle$, with labels from $ProgAct$, where  $\AllPrograms$ is all possible program states, $P_0$ is the initial state of the program,  
and the transition relation ${\rightarrow} \subseteq (\AllPrograms \times \ProgAct \times \AllPrograms) $ is deadlock-free, and satisfies the following determinacy property: for all states $P$, if $ {P} \Sstep{a}{} {P'} $ and ${P} \Sstep{b}{}{P''}$ then
\begin{enumerate}
    \item if $a=b$ then $\Program{P'}=\Program{P''}$, 
    \item if $a$ is not a query then  $a=b$, 
    \item if $a= \QueryAct(\tvar, f, \QueryVec{\epsilon})\ ?\ \vect{v} $ then $b=\QueryAct(\tvar, f, \QueryVec{\epsilon})\ ?\ \vect{u} $ for some $\vect{u}$, and for all actions $c$ of the form  
$\QueryAct(\tvar, f, \QueryVec{\epsilon})\ ?\ \vect{w}$ there exists a state $P_c$ such that 
${P} \Sstep{c}{}  {P_c} $.
\end{enumerate}
\end{definition}
The conditions on client programs are mild.  Deadlock (i.e. termination) freedom simplifies reasoning; a program that terminates in the conventional sense can be modelled by adding a transition $P  \Sstep{\tau}{} P$ for all terminated states $P$.   Query transitions model both the query sent and the result received. Since we are modelling message passing using just transition labels, the condition on queries states that the program must be able to accept any result from a given query. Modulo the results returned by a query, the conditions require the program to be deterministic. This is a technical simplification which (we believe) does not restrict the power of the attacker. 

\paragraph{Remark: Implicit parameters}
We will prove that Featherweight PINQ provides differential privacy for any client program. To avoid excessive parametrisation of subsequent definitions, in what follows we will fix some arbitrary client program $\langle \AllPrograms, {\rightarrow}, P_0 \rangle$ and some arbitrary initial budget $\epsilon$ and make definitions relative to these.



\section{Featherweight PINQ}
\label{subsec:sysmodel}
In this section we turn to the model of the internals of PINQ, and the overall semantics of the system. We begin by describing the components of the protected system, and then give the overall model of Featherweight PINQ by giving a probabilistic semantics (as a probabilistic labelled transition system) to the combination of a client program and a protected system. 


\subsection{The Protected System}\label{subsec:protected}
\paragraph{Global Privacy Budget} The first component of the protected system is the global privacy budget. This is a non-negative real number representing the remaining privacy budget. The idea is that if we begin with initial budget $b$ then Featherweight PINQ will enforce $b$-differential privacy. The global budget is decremented as queries are computed, and queries are denied if they would cause the budget to become negative.  In PINQ the budget is associated with a given data source. In our model we assume that there is only one data source, and hence only one budget. Further, PINQ allows the budget to be divided up and passed down to subcomputations. This does not fundamentally change the expressiveness of PINQ since, as we show later, we are free to extend Featherweight PINQ with the ability to query the global budget directly. Thus any particular strategy for dividing the global budget between subcomputations can be easily programmed.



\paragraph{The Table Environment}
The other data component of the protected system is the table environment, which maps each table variable to the table it denotes, together with a record of the \emph{scaling factor}, which is a measure of the \emph{stability} of the table relative to the initial data set. We define this precisely below. 
%
%
Formally we define a table as power-set of records, $\mathcal{P}(\RecordSet)$, a protected table is a pair of a table with its scaling factor:
\begin{align*}
  \protectedTable & \eqdef  \AllTables \times \mathds{N}  
\end{align*}

\subsection{The Featherweight PINQ Transition System} 
Featherweight PINQ is defined by combining a client program with the protected system to form the states of a probabilistic transition system.
 \begin{definition}[Featherweight PINQ States]
The \emph{states} (otherwise known as \emph{configurations}) of Featherweight PINQ, ranged over by $\Conf$, $\Conf'$ etc., are triples of the form 
$\Configuration{P}{E}{B}$ where $P$ is a client program state, $E \in \TVar \rightarrow \protectedTable$ is the table environment, and $B \in \mathds{R}^{+}$ is the global budget.
\end{definition}
There is a family of possible initial states, indexed by the distinguished input table, and the initial budget. We define these by assuming the existence of a distinguished table variable, $\varName{input}$, which we initialise with the input table, while all other table variables are initialised with the empty table: 
\begin{definition}[Initial configuration]\label{def:initconfig} 
\[   \InitConf(T,\Budget) \eqdef 
    \Configuration
        {P_0}
        {\Env{E}_T}   
        {\Budget}\ \text{where}\ \Env{E}_T(\tvar) \triangleq 
		\begin{cases}
		    (T,1)\ \text{if}\ \tvar = \varName{input}\\
		    (\{\},0)\ \text{otherwise.}
		\end{cases}
\]
\end{definition}
The operational semantics of featherweight PINQ can now be given:
\begin{definition}[Semantics]
The operational semantics of configurations is given by a probabilistic labelled transition relation with transitions of the form
$\Conf \Sstep{a}{p} \Conf'$ where $a \in \Act \eqdef \{\tau,\bot\} \Union \bigcup_{n\in \Nat}\Val^n$, 
and (probability) $p \in [0,1]$.  The definition is given by cases 
in Figure~\ref{fig:semantics}.
\end{definition}

\newcommand{\RULEQUERY}[1][Query ]
{
 \inference[#1]{
                    {P} \Sstep{
                    \QueryAct(\tvar,f,\vect{Q},\epsilon)\ ?\ \vect{v}
                    }{} {P'} 
                }{
                    {\Configuration{\Program{P}}{\Env{E}}{\Budget}}
                    \Sstep{\vect{v}}{p}
                    {\Configuration{\Program{P'}}{\Env{E}}{\Budget- s \cdot  \epsilon}}
                } 
\text{where}
\begin{cases}
\Env{E}(\tvar)= (T,s), \quad \epsilon \cdot s \leq B 
\\
\codomain(f) = \Set{1\upto n} \quad \vect{v} \in \Val^n
\\
T_i = \{s \mid s \in T,  f(s)= i \}, i \in \Set{1\upto n}
\\
p = \prod_{i=1}^n \Pr[\Query{i}(\epsilon, T_i) = v_i ] 
\end{cases}
}

\newcommand{\RULEQUERYEXCEPTION}[1][Query\!$_\bot$]
{
 \inference[#1]{
                    {P} \Sstep{
		    \QueryAct(\tvar,f,\vect{Q},\epsilon)\ ?\ \bot 
                    }{} {P'} 
                }{
                    {\Configuration{\Program{P}}{\Env{E}}{\Budget}}
                    \Sstep{\bot}{1}
                    {\Configuration{\Program{P'}}{\Env{E}}{\Budget}}
                } 
\text{where} 
\begin{cases}
\Env{E}(\tvar) = (T, s) \\
\epsilon \cdot s > B  
\end{cases}
}

\newcommand{\RULEASSIGN}[1][Assign]
{
\inference[#1]{
                    {P} \Sstep{\tvar := F(\tvar_1, \ldots, \tvar_n)}{} {P'}
                }{
                    {\Configuration{\Program{P}}{\Env{E}}{\Budget}}
                        \Sstep{\Silent}{1}
                    {\Configuration{\Program{P'}}{\Env{E}[\tvar \mapsto (T',s)]}{\Budget}}
                } 
\ \text{where}
\begin{cases}
    \Env{E}(\tvar_i) = (T_i, s_i), i \in \Set{1\upto n} \\
    \SensitivityFunc(F) = (c_1\upto c_n) \\
    s = \sum_{i=1}^n c_i \times s_i  \\
    T' = \llbracket F \rrbracket(T_1\upto T_n)  \\
\end{cases}
}

\begin{figure*}[ht]
\centering
\setpremisesend{0pt}
\[
\begin{array}{l}
\inference[Silent]{
                    {P} \Sstep{\Silent}{} {P'}
                }{
                    {\Configuration{\Program{P}}{\Env{E}}{\Budget}}
                        \Sstep{\Silent}{1}
                    {\Configuration{\Program{P'}}{\Env{E}}{\Budget}}
                }

\\
\RULEASSIGN
\\[3ex]
\RULEQUERYEXCEPTION
\\[3ex]
\RULEQUERY
\end{array}
\]
  \caption{Operational semantics}
  \label{fig:semantics}
\end{figure*}

We note at this point that some of the primitives have not yet been defined (e.g. $\opName{stability}$ in the Assign rule), and that the rules of the system do not, \emph{a priori}, define a probabilistic transition system. We will elaborate these points in what follows. We begin by explaining the rules in turn. 
\paragraph{Assign}
When a program issues an assignment command 
$\tvar := F(\tvar_1,\ldots,\tvar_n)$,
the value of the stored table for $\tvar$ is updated in the obvious way. We must also record the scaling factor of the table thus computed. The scaling factor is computed from the scaling factors of the tables for $\tvar_1, \ldots, \tvar_n$, and the \emph{stability} of the transformation $f$. We assume a mapping $\llbracket \cdot \rrbracket$ from formal function identifiers $F$ to the actual table transformation functions $\llbracket F \rrbracket$ of corresponding arity. 
\begin{definition}
\label{definition:stability}
A table transformation $f$ of arity $n$ has stability $(c_1, \ldots, c_n)$ if for all $i \in \Set{1,\ldots,n}$, we have 
\[
\Length{f(T_1, \ldots, T_i, \ldots,  T_n)\ominus f(T_1, \ldots, T'_i, \ldots T_n)} \leq c_i \times \Length{T_i \ominus T'_i}
\]
\end{definition}
This is the n-ary generalisation of McSherry's definition \cite{mcsherry2009privacy}, and bounds the size change in a result in terms of the size change of its argument. This is made more explicit in the following:
\begin{lemma} 
\label{lemma:scalingfactor}
		If $f$ has stability $(c_1, \ldots, c_n)$ then $\Length{ 
			f(T_1, \ldots , T_n) \ominus f(T'_1, \ldots ,T'_n)
		}
		\leq  \sum_{i}^{n}(c_i \times |T_i \ominus T'_i|) $
\end{lemma}
Note that not all functions have a finite stability. An example of this is the database join operation (essentially the cartesian product); adding one new element to one argument will add $k$ new elements to the result, where $k$ is the size of the other argument.  Thus there is no static bound on the number of elements that may be added.  Thus PINQ (and hence Featherweight PINQ) supports only transformation operations which have a finite stability. Table~\ref{table:functionstability} illustrates the stability of some of the transformations that are introduced in PINQ. The variant of the join operation, $\mathrm{Join}^{*}$ deterministically  produces bounded numbers of join elements. For the purpose of this paper we do not need to be specific about the transformations.  We simply assume the existence of a function $\opName{stability}$ which soundly returns the stability of a function identifier, i.e., if $\opName{stability}(F) = (c_1, \ldots, c_n)$ then $\llbracket F \rrbracket$ has stability $(c_1, \ldots, c_n)$.

The transition rule for assignment in featherweight PINQ is thus
\[
\RULEASSIGN[]
\]
The label on the rule $\tau$ says that nothing (other than computational progress) is observable from the execution of this computation step.  The subscript $1$ is the probability with which this step occurs.
\begin{table}[ht]
\begin{center}
\begin{tabular}{ l | c }
\label{table:functionstability}
  Transformation & Stability \\
\hline
  Select($T$, $maper$) & (1) \\
  Where($T$, $predicate$) & (1) \\
  GroupBy($T_1$, $keyselector$) & (2)  \\
  Join*($T_1$,$T_2$, $n$, $m$, $keyselector_1$, $keyselector_2$) & (n,m)  \\
  Intersect($T_1$,$T_2$) & (1,1) \\
  Union($T_1$,$T_2$) & (1,1)  \\
  Partition($T$, $keyselector$, $keysList$) & (1) \\
\end{tabular}
\end{center}
\caption{Transformation stability}
\end{table}



\begin{figure}[bth]
\begin{minipage}[t][][b]{0.6\textwidth}
\centering   
\vspace{60pt} 
\includegraphics[trim=0cm 10cm 3cm 8cm, 
 width=\textwidth
 ]{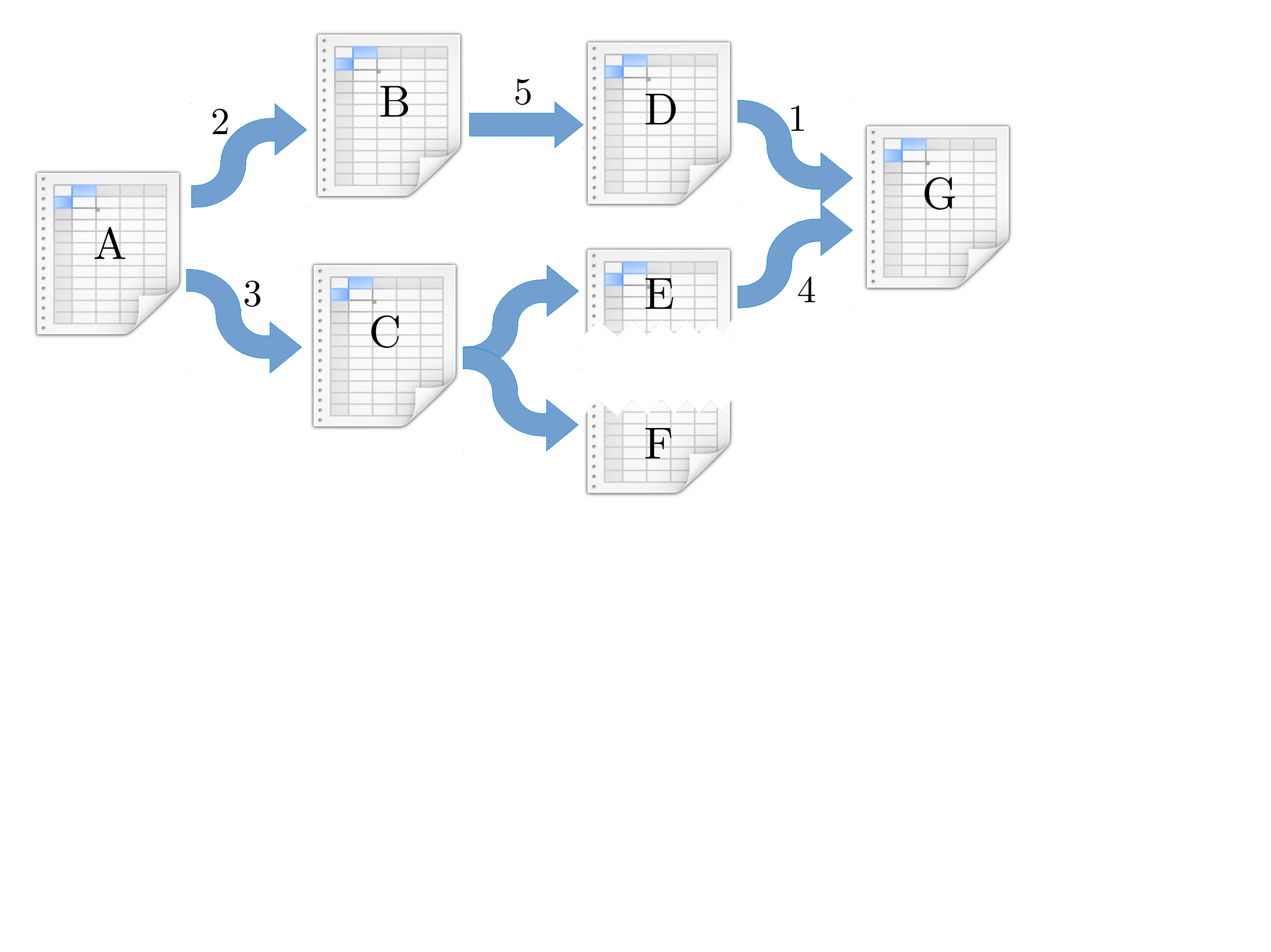}
\caption{Transformations}
\label{fig:env}
\end{minipage}
\begin{minipage}[t][][b]{0.4\textwidth}
\centering
\begin{tabular}{l|c|@{\;}l}
\small & s & Calculation \\
\hline
A     & 1  &  Input table \\
B     & 2  & $s(A) \times 2 $\\
C     & 1  & $s(A) \times 3 $\\
D     & 10 & $s(B) \times 5$ \\
E     & 3  & $s(C)$  \\
F     & 3  & $s(C)$  \\  
G     & 22 & $s(D) \times 1 + s(E) \times 4$ \\
\end{tabular}
\caption{Scaling factors (s)}
\label{scalingfactor}
\end{minipage}
\end{figure}

\paragraph{Understanding the scaling factor}
Here we provide more intuition about the scaling factor calculations, and explain some differences between the PINQ implementation and the Featherweight PINQ model. 
%
%
%
As an example, suppose we have a computation of a series of tables $A$--$G$ depicted in Figure~\ref{fig:env}.  

The figure represents a PINQ computation involving three unary transformations (producing $B$, $C$, and $D$), one binary transformation producing $G$, and one partition operation (splitting $C$ into $E$ and $F$). We have labelled the transformation arcs with the stability constants of the respective transformations. What is the privacy cost of an $\epsilon$ differentially private query applied to, say, table $D$? Since $D$ is the result of two transformations on the input data, the privacy cost is higher than just $\epsilon$. The product of the sensitivities on the path from $D$ backwards to the input $A$ provide the \emph{scaling factor} for $\epsilon$. In this case the scaling factor for a query on $D$ is $10$. The remaining scaling factors are summarised in the table in Figure~\ref{scalingfactor}.



The scaling factor is the stability of that specific table; it bounds the maximum possible change of the table as a result of a change in the input dataset, assuming that it was produced using the same sequence of operations. The scaling factor is computed from the stabilities of transformations that produced it.
The scaling factor for each protected table (except input table which has the scaling factor one) is computed compositionally using the scaling factors ($s_i$) of all the arguments  and the sensitivities of corresponding transformation arguments ($c_i$) using the following formula:    
$ \mbox{s}_A= \sum_{i \in parent(A)}{\mbox{c}_i \times \mbox{s}_i}$



Figures~\ref{fig:env} and \ref{scalingfactor} allow us to explain two key differences between PINQ and our model:  
\begin{enumerate}
\item In PINQ, the tree structure depicted in the figure is
  represented explicitly, and scaling factors are calculated lazily:
  at the point where a query with accuracy $\epsilon$ is made on a
  table it is necessary to calculate its scaling factor $s$ in order
  to determine the privacy cost $s.\epsilon$. To do this the tree is
  traversed from the query at the leaf back to the root, calculating
  the scaling factor along the way. At the root the total privacy cost is then
  known and deducted from the budget (providing the budget is
  sufficient). In Featherweight PINQ the scaling factors of each table
  are computed eagerly, so the tree structure is not traversed.
\item   In Featherweight PINQ we restrict the partition operation to
  the leaves of the tree, and combine it with the application of primitive queries to partitions.
\end{enumerate}
The consequence of these two simplifications is that we do not need to represent the PINQ computation tree at all -- all computations are made locally at the point at which a table is produced or queried.

\paragraph{Queries} 
Parallel queries were described in detail in the previous section. When a program issues a query is it represented as a parallel query and a possible result -- i.e. we model the query and the returned result as a single step.  There are two cases to consider, according to whether the budget is sufficient or not. If the queried table $T$ has scaling factor $s$ then the cost of an $\epsilon$ query is $s\times\epsilon$. If this is greater than the current global budget then the result is the exceptional value $\bot$.  This value is the observable result of the query, and it occurs with probability $1$. On the other hand, if the budget is sufficient, then the vector of query results $\vect{v}$ is returned with probability  $p = \prod_{i} \Pr[\Query{i}(\epsilon, T_i) = v_i ]$ where $T_i$ is the $i$th partition of $T$. Note that $p$ is indeed a probability, since the component queries are independent.



\section{Differential Privacy for Featherweight PINQ}
\label{sec:dp}
In this section we prove that Featherweight PINQ is differentially
private. We begin by recapping the goals of differential privacy,
before showing how to specialise the definition to Featherweight
PINQ. Doing this entails building a trace semantics for Featherweight
PINQ. 


Differential privacy, guarantees that a data query mechanism (abstractly, a randomized algorithm) behaves similarly on
similar input databases. This ``similarity'' is a quantitative
measure $\epsilon$ 
on the difference in the information obtained from any data set with
or without any individual.  When this difference is small, the
presence or absence of the individual in the data set is difficult to
ascertain. 


%

\begin{definition}
\label{dpdef}
Mechanism f provides $\epsilon$-differential privacy if for any two datasets $A$ and $B$ that differ in one record  ($ \mid A \ominus B \mid =1 $), and for any two possible outcome $f(A)$ and $f(B)$, the following inequalities holds : 
$
{\mathrm e}^{- \epsilon}
\leq
\frac 
	{ \Probability{(f(A) \in \RangeS)}}
	{ \Probability{(f(B) \in \RangeS)}}
\leq 
{\mathrm e}^{\epsilon}
$
\end{definition}

In this definition, $\RangeS$ is subset of the range of outcomes for $f$ ($\RangeS \subseteq \mathrm{Range}(f)$) and for similarity of outcomes we use the ratio between the probabilities of observing outcomes $\Probability{(f(datasets) \in  \RangeS)}$ when the analyses are executed on any two similar datasets $A$ and $B$. Finally for similarity of datasets hamming distance is used as a metric. In this work we assume that the primitive query mechanisms (and thus Featherweight PINQ) provide answers over a discrete probability distribution, so that it is sufficient to consider $\RangeS$ to be a singleton set.

\subsection{Trace semantics}\label{sec:traceSemantics}
The first step to instantiating the definition of differential privacy to Featherweight PINQ is to be able to view Featherweight PINQ as defining a probabilistic function. In fact each client program gives rise to a family of probabilistic functions, one for each length of computation that is observed. This is given by building a \emph{trace semantics} on top of the transition system for Featherweight PINQ. 

The semantics of Featherweight PINQ is a probabilistic labelled
transition system of the simplest kind: for each configuration $\Conf$, 
the sum of all probabilities of all transitions of $\Conf$ is equal to
$1$. The system is also deterministic, in the sense that if ${\Conf}
\Sstep{a}{p_1} {\Conf_1}$ and  ${\Conf} \Sstep{a}{p_2} {\Conf_2}$ then
$p_1 = p_2$ and $\Conf_1 = \Conf_2$. This makes it particularly easy
to lift the probabilistic transition system from single actions to
traces of actions:
\begin{definition}[Trace semantics]\label{def:tracesemantic}
  Define the trace transitions  ${\Rightarrow} \subseteq \Configs
  \times \Act^{*} \times [0,1] \times \Configs$ inductively as
  follows: (i) ${\Conf}
            \Mstep {\EmptyTrace}{1}
            {\Conf}$ where $\EmptyTrace \in \Act^{*}$ is the empty
            trace, and (ii) if ${\Conf} \Sstep{a}{p} {\Conf'}$ and
            ${\Conf'} \Mstep{\trace}{q}  {\Conf''}$ then 
            ${\Conf}
                    \Mstep{a\trace}{p.q} 
                    {\Conf''}$
\end{definition}
Traces inherit determinacy from the single transitions: 
\begin{proposition}[Traces are Deterministic]
If $\Conf \Mstep{\trace}{p_1} \Conf_1 $ and  $\Conf \Mstep{\trace}{p_2} \Conf_2$ then $p_1 = p_2$ and $\Conf_1 = \Conf_2$
\end{proposition}
This follows by an easy induction on the trace, using the fact that the single step transitions are similarly deterministic. 
\begin{lemma}[Traces are Probabilistic]\label{lemma:probabilistic}
Define 
\[ \mu(\Conf, t) \eqdef \begin{cases}
                        p & \text{if $\Conf \Mstep{t}{p}{} \Conf'$} \\
                        0 & \text{otherwise.}
                       \end{cases}
\]
For all configurations $\Conf$, and all $n >0$, 
\[
\sum_{t \in Act^n} \! \mu(\Conf, t) \; = 1
\]
where $Act^n$ is the set of traces of length $n$.
\end{lemma}
The proof is a simple induction on $n$, using the proposition above. 
%
The lemma says that whenever  $\Conf \Mstep{\trace}{p}$, then $p$ is the probability that you see trace $\trace$ after having observed $\Size{t}$ steps of the computation of $\Conf$. We will thus refer to the probability of a given trace to mean the probability of producing that trace from the given configuration among all traces of the same length. 
We denote this by writing $\Probability{\Conf \Mstep{\trace}{}} = p$ when $\Conf \Mstep{\trace}{p}$.

\paragraph{Differential Privacy for Traces}
We are now in a position to specialise the definition of differential privacy for Featherweight PINQ. How can we view Featherweight PINQ as a probabilistic function? The probabilistic function is determined by the client program (which we have kept implicit but unconstrained), the initial budget $\epsilon$, and the length of trace $n$ that is observed for any combination of these we define the function which maps a table $T$ to trace $t$ of length $n$ with probability $p$ precisely when $\Probability{\InitConf(T ,\epsilon) \Mstep{t}{}} = p$. 

The instantiation of the differential privacy condition to Featherweight PINQ is thus: 
\[
\label{tracedp}
\ForAll{t, T, T', \epsilon}
{\text{if $ |T \ominus T'| = 1 $ then }
{\mathrm e}^{- \epsilon}
\leq
\frac 
	{ \Probability{\InitConf(T ,\epsilon) \Mstep{t}{}}}
	{ \Probability{\InitConf(T',\epsilon) \Mstep{t}  }}
\leq 
{\mathrm e}^{\epsilon} 
}
\]


Towards a proof of this property we introduce some notation to reflect
key invariants between the pairs of computations (for $T$ and $T'$
respectively).
\begin{definition}[Similarity]
\label{def:mySim}\label{def:simenv}
We define similarity relations $\sim$ between tables, environments, and configurations as follows:
\begin{itemize}
\item For tables $T$ and $T'$, and $s \in \Nat$ define $T \sim_s T'$
  (``$T$ is $s$-similar to $T'$'') if and only if $\Length{T
    \ominus T'} \leq s$.
\item For protected environments $E$ and $E'$, define $\mySim{E}{E'}{}$ if
  and only if for all $\tvar$, if $E(\tvar) = (T,s)$ and $E'(\tvar) =
  (T',s')$ then $s = s'$ and $T \sim_s T'$.
\item For configurations, define
  $\mySim{\Configuration{P}{E}{B}}{\Configuration{P'}{E'}{B'}}{}$ if
  and only if $P = P'$, $\mySim{E}{E'}{}$ and  $B =  B'$.
\end{itemize}
\end{definition} 



The configuration similarity relation captures the key invariant
between the two computations in our proof of differential
privacy. First we need to show that the invariant is established for
the initial configurations: 
\begin{lemma} \label{lemma:init-similarity} 
   If $T \sim_1 T'$ then $\InitConf(T, \Budget) \sim \InitConf(T', \Budget)$.
\end{lemma}
This follows easily form the definition of the initial
configuration. Now the main theorem shows that this is maintained
throughout the computation:
\begin{theorem} \label{theorem:indistinguishable} 
   If $T \sim_1 T'$ and $\InitConf(T, \Budget) \Mstep{\trace}{p} \Conf$, then $\InitConf(T',\Budget) \Mstep{\trace}{q} \Conf'$ where 
$\Conf \rSim{} \Conf'$  and 
$p \leq q . \exp(B - \epsilon)$  for some $\epsilon \leq \Budget$.
\end{theorem}
The proof, an induction over the length of the trace, is given in Appendix~\ref{prooflemma}.

\begin{corollary}[$\Budget$-differential privacy]\label{theorem:dp}
If $T \sim_1 T'$  and
$\Probability{{\InitConf(T,\Budget)}\Mstep{\trace}{}} = p$
then $\Probability{{\InitConf(T', \Budget)} \Mstep{\trace}{}} = q$ for some $q$ such that 
$p \leq q \cdot \exp(\Budget)$.
\end{corollary}

\section{Related Work}
The approach described in this paper owes much to the model used in the formalisation developed in our recent work on \emph{personalised differential privacy} \cite{PersonalisedDifferentialPrivacy}. The idea to model the client program as an abstract labelled transition system comes from that work. That work also shows how dynamic inputs can be handled without major difficulties.

 The closest other prior work is developed by Tschantz et al \cite{tschantz2011formal}.
Their work introduces a way to model interactive query mechanisms  as a probabilistic automata, and develop bisimulation-based proof techniques for reasoning about the differential privacy of such systems. As a running example they consider a system ``similar to PINQ'', and use it to demonstrate their proof techniques. From our perspective their system is significantly different from PINQ in an number of ways: (i) it does not model the transformation of data at all, but only queries on unmodified input data, (ii) it models a system with a bounded amount of memory, and implements a mechanism which deletes data after it has been used for a fixed number of queries (neither of which relate to the implementation of PINQ). 
Regarding the proof techniques developed in \cite{tschantz2011formal}, as previously noted in \cite{PersonalisedDifferentialPrivacy}, a key difference between our formalisation and theirs is that they model a passive system which responds to external queries from the environment. In contrast, our model includes the adaptive adversary (the client program) as an explicit part of the configuration. In information-flow security (to which differential privacy is related) this difference in attacker models can be significant \cite{Wittbold:Johnson:Information}. However it may be possible to prove that the passive model of \cite{tschantz2011formal} is sound for the active model described here (c.f. a similar result for interactive noninterference \cite{Clark:Hunt:FAST08}).

Haeberlen et al \cite{haeberlen2011differential} point out a number of flaws an covert channels in the PINQ system. This may seem at odds with our claims for the soundness of PINQ, but in fact all the flaws described are either covert timing channels (which we do not attempt to model), flaws in PINQ's implementations of encapsulation, or failure to prevent unwanted side-effects, or combinations of these.  
 Following this analysis, Haeberlen et al introduce a completely different  approach to programming with differential privacy (an approach further developed and refined in \cite{reed2010distance} \cite{gaboardi-2013-dfuzz}) based on statically tracking 
sensitivity through sensitivity-types. This non-interactive approach is rigorously formalised and proven to provide differential privacy.

Barthe et al \cite{barthe2012probabilistic} introduce a relational Hoare-logic for reasoning formally about the differential privacy of algorithms. They include theorems relating to sequential and parallel composition of queries in the style of those stated by McSherry \cite{mcsherry2009privacy}. Unlike the present work, \cite{barthe2012probabilistic} does not rely on differentially private primitives, but is able to prove differential privacy from first principles.


\section{Limitation and Extension}
\label{sec:limitation}
In this section we discuss what we see as the main limitations of  Featherweight PINQ in relation to the PINQ system. We also discuss some easy extensions that become apparent from the proof of differential privacy. 
\paragraph{Parallel Queries}
The form of parallel query that we model matches the informal description in \cite{mcsherry2009privacy}, but is not as general as the construct found in the implementation.  We believe that this is the main shortcoming in the Featherweight PINQ model, as more general form is interesting, and thus its correctness is not immediately obvious. (Whether the shortcoming has any practical significance in the way one might write programs is less clear.) The difference was described in Section~\ref{subsec:sysmodel} in connection with Figure~\ref{fig:env}, which depicts a partition operation which is not supported by Featherweight PINQ since it is not immediately followed by queries on the partitions.  In fact the queries in PINQ need not be parallel at all, but can be adaptive (i.e., a query on one partition can be used to influence the choice of query on other partitions). This change is not easily supported by a small change to our model since it does not seem to be implementable using Featherweight PINQ's simple history-free use of explicit scaling factors. A proof of differential privacy for a more general protected system model encompassing this is left for future work. 

\paragraph{Extensions to the PINQ API}
We mention one extension to PINQ that emerges from the details of the correctness proof. In PINQ, the budget and the actual privacy cost of executing an $\epsilon$ differentially private query on some intermediate table is not directly visible to the program: 
\begin{quote}
  \it ``An analyst using PINQ is uncertain whether any request
will be accepted or rejected, and must simply hope that the
underlying PINQAgents accept all of their access requests.'' \rm \cite{mcsherry2009privacy}(\S{3.6})
\end{quote}
 Recall that the key invariant that relates the two runs of the systems on neighbouring data sets (Definition~\ref{def:mySim}) states that the budgets and the scaling factors in the respective environments are equal.  This means that they contain \emph{no} information about the sensitive data. This, in turn, means that we can freely permit the program to query them.  This would allow the analyst to calculate the cost of queries and to make accuracy decisions relative to the current privacy budget.

Here we briefly outline this extension. 
We add two new actions to the set of
program actions $\ProgAct$, namely a query on the sensitivity of a
table variable of the form $\tvar\ ?\ s$, where $s \in \Nat$, and a
query on the global budget, $ \opName{budget} \ ?\ v$ where $r \in
\Real^{\geq 0}$. The transition rules are given in
Figure~\ref{fig:extension}. 
\begin{figure*}[h]
\centering
$
\begin{array}{l}
\inference[Query sensitivity]{
                    {P} \Sstep{ \tvar ? s}{} {P'}
                }{
                    {\Configuration{\Program{P}}{\Env{E}}{\Budget}}
                        \Sstep{\tau}{1}
                    {\Configuration{\Program{P'}}{\Env{E}}{\Budget}}
                } 
		\text{where}\ \Env{E}(\tvar) = (T, s) 
\\[3ex]
\inference[Query budget]{
                    {P} \Sstep{ \opName{budget} ? \Budget }{} {P'}
                }{
                    {\Configuration{\Program{P}}{\Env{E}}{\Budget}}
                        \Sstep{\tau}{1}
                    {\Configuration{\Program{P'}}{\Env{E}}{\Budget}}
                }
\end{array}
$
  \caption{Budget- and Scaling-Factor-Query}
\label{fig:extension}
\end{figure*}


\section{Conclusion}
We started by presenting some shortcomings(gaps) between the theory of differential privacy and the implementation of PINQ framework. To verify privacy assurance of analysis written in PINQ framework and to address the mentioned concerns, we introduced an idealised model for the implementation of PINQ. 
In the model, only PINQ's internal implementation has direct access to the sensitive data. An analysis written in this framework has indirect access to the protected system by calling some limited well defined/crafted interface APIs. In addition to the standard PINQ APIs, we extended the model with our own proposed APIs responsible to retrieve scaling factor and the budget from the protected environment. Furthermore we instantiated the definition of differential privacy to prove any analysis constructed in this setting and its communications with protected system would not violate the privacy guarantee promised by PINQ.

We believe that our model (and our general approach to modelling such systems) could be of benefit to formalise emerging variants on the PINQ framework, such as wPINQ \cite{davide2012calibratingdata}, or Streaming PINQ  \cite{StreamingPINQ}.

\bibliographystyle{plain}
\bibliography{Bibliography,Bibliography2}

\appendix
 
\section{Proof of Theorem \ref{theorem:indistinguishable}}
\begin{proof} 
\label{prooflemma}
Assume $\InitConf(T, \Budget) \Mstep{\trace}{p} \Conf$. We proceed by induction on the length of the trace $\trace$, and by cases according to the last step of the trace. 

\proofcase[Base case]{$\trace = \EmptyTrace$} In this case $p=q=1$ and $\Conf = \InitConf(T, \Budget)$ and $\Conf' = \InitConf(T', \Budget)$. So $\epsilon=\epsilon'=0$ and $\Conf \rSim{} \Conf'$.
\\
\proofcase[Inductive step]{$\trace = \trace_1 a$}
Suppose that $\InitConf(T, \Budget) \Mstep{\trace_1}{p_1} \Configuration{P_1}{E_1}{B_1} \Sstep{a}{p_2} \Configuration{P}{E}{B} = \Conf$, and hence that $p = p_1p_2$.

The induction hypothesis gives us $q_1$, $P_1$, $E_1'$ and $\epsilon_1$ such that
\begin{align}
  \label{eq:IH1}
  & \InitConf(T', \Budget) \Mstep{\trace_1}{q_1} \Configuration{P_1}{E_1'}{B_1} 
\\
\label{eq:IHE}
  & E_1 \rSim{}  E_1'
\\ 
\label{eq:IHp} & p_1 \leq q_1 . \exp(B - \epsilon_1)
\end{align} 
by cases that is applied to the rule as the last transition 
($\Configuration{P_1}{E_1}{B_1} \Sstep{a}{p_2} \Configuration{P}{E}{B} $) 
we have $p_2 = 1$ except for query execution and that $\Configuration{P_1}{E_1'}{B_1} \Sstep{a}{1} \Conf'$
for some $\Conf'$. In those cases it  follows that 
$p \leq q \cdot \exp(B - \epsilon)$ by taking $\epsilon =
\epsilon_1$ and using (\ref{eq:IHp}).

\proofcase[Case 1]{Silent} In this case $a = \Silent$ and $P_1 \Pstep{\Silent} P$. 
\[
\Configuration{P_1}{E_1}{B_1} \Sstep{\Silent}{1} \Conf = \Configuration{P_1}{E_1}{B_1}
\] 
\[
\Configuration{P_1}{E_1'}{B_1} \Sstep{\Silent}{1} \Conf' = \Configuration{P_1}{E_1'}{B_1}
\]
It follows directly from (\ref{eq:IHE}) that $\Conf \rSim{ } \Conf'$. 

\proofcase[Case 2]{Assign} Here 
$P_1 \Pstep{t:= \apply{F}{t_1 , \dots , t_n}} P$, 
and so we have
 \begin{align*}
\Conf &= \Configuration{P}{E_1[\tvar \mapsto (T,s)]}{B_1}
\\
\Conf'&= \Configuration{P}{E_1'[\tvar \mapsto (T',s) ]}{B_1}
\end{align*}
where for $i \in (1, \ldots, n)$
\begin{align*}
E_1(\tvar_i)=(T_i, s_i) 
\\
E_1'(\tvar_i)=(T_i', s_i')
\\
\SensitivityFunc(F) = (c_1, \ldots ,c_n)
\\
s = \sum_{\sum_{i}^{n}} c_i \times s_i
\\
T= \apply{\llbracket F \rrbracket}{T_1, \ldots , T_n}
\\
T'= \apply{\llbracket F \rrbracket}{T_1', \ldots , T_n'}
\end{align*}
From~(\ref{eq:IHE}) we have  
$E_1(\tvar_i) \rSim{} E_1'(\tvar_i)$ which means $s_i = s_i'$ and $ T_i \sim_{s_i}  T_i' $.
Using similarity definition and Lemma~\ref{lemma:scalingfactor} we have $T  \sim_s  T' $ and hence we have $\Conf \rSim{} \Conf'$.


\proofcase[Case 3]{Query}
The result of query execution depends on the remained budget and the sensitivity of the table that the query is executed on. If privacy budget is insufficient an exception is thrown to inform the program about the shortage of budget, otherwise each query in the list of queries will be executed on its corresponding partition and the result of execution is returned as a list of values, $\vect{v}$. 
\\
\proofcase[Case 3.1]{Query(run out of budget)} Here we have a rule instance of the form:  
\[ \RULEQUERYEXCEPTION  \]
In this case $\Conf \rSim{} \Conf'$ and is similar to silent case.  

\proofcase[Case 3.2]{Query}
Similarly we have a rule instance of the form:  
\[ \RULEQUERY \]
Hence we have a transition : $ \Configuration{P_1}{E_1}{B_1} \Sstep{\vect{v}}{p_2} \Conf = \Configuration{P}{E}{B} $ and the analogous transition : $\Configuration{P_1}{E_1'}{B_1} \Sstep{\vect{v}}{q_2}  \Conf' = \Configuration{P}{E'}{B}$. 
$\epsilon = \epsilon_1 + (s \cdot \epsilon_2)$ is the value needed for theorem \ref{theorem:indistinguishable}.
\\For parallel queries on disjoint set we have the following equation:
\[
\Pr[{P_1} \Sstep{ \QueryAct(\tvar,f,\vect{Q},\epsilon)\ ?\ \vect{v}  }{}{P}]= \prod_{i=1}^n{\Pr[\Query{i}(s \cdot \epsilon_2, T_i)}=v_i]
\]
Here we need to show that the following inequality is valid:
\[
\prod_{i=1}^n{\Pr[\Query{i}(s \cdot \epsilon_2, T_i)}=v_i]
\leq 
\prod_{i=1}^n{\Pr[\Query{i}(s \cdot \epsilon_2 , T'_i)}=v'_i] \times \prod_{i=1}^n \exp(\epsilon_2 \times \mid T_i - T'_i\mid) 
\]
From $\sum_{i=1}^n( \mid T_i - T'_i\mid) = s $, we have $\prod_{i=1}^n \exp(\epsilon_2 \times \mid T_i - T'_i\mid) \leq \exp(\epsilon_2 \times s) $
which we conclude:
\[
\prod_{i=1}^n{\Pr[\Query{i}(s \cdot \epsilon_2, T_i)}=v_i]
\leq
\prod_{i=1}^n{\Pr[\Query{i}(s \cdot \epsilon_2, T'_i)}=v'_i] \times \exp(\epsilon_2 \cdot s)
\]
These parallel queries provide $(s\cdot\epsilon)$-differential privacy which means:
\[
p_2 \leq q_2 \cdot \exp(\epsilon_2 \cdot s ) 
\]
Multiplying two sides of the previous inequality with (\ref{eq:IHp}) we get:
\begin{align*}
p_1 \cdot p_2 & \leq q_1 \cdot q_2 \cdot
\exp(\epsilon_1) \cdot  
\exp(\epsilon_2 \cdot s)
\end{align*}
Knowing $\Budget_1 = \Budget - \epsilon_1$ result in choosing $\epsilon$ to be $\epsilon = \Budget - \epsilon_1 - (\epsilon_2 \cdot s)$. 
Finally it is easy to see $\Conf \rSim{} \Conf'$ as the proper reduction in the global budget is the only change in the configuration.
\end{proof}

 
\end{document}